\documentstyle[12pt]{article}

\def\be{\begin{equation}}
\def\ee{\end{equation}}
\def\ben{\begin{displaymath}}
\def\een{\end{displaymath}}
\def\ba{\begin{array}{c}}
\def\bal{\begin{array}{l}}
\def\ea{\end{array}}

\begin{document}

.

\vspace{.35cm}

 \begin{center}
 {\Large \bf
 Construction of a unique metric in quasi-Hermitian quantum mechanics:
 non-existence of the charge operator in a $2\times 2$ matrix model
  }\end{center}

\vspace{3mm}

 \begin{center}

 {\bf Miloslav Znojil}

 \vspace{2mm}
 {\it
\'{U}stav jadern\'e fyziky AV \v{C}R,

250 68 \v{R}e\v{z}, Czech Republic

{e-mail: znojil@ujf.cas.cz}
}

\vspace{2mm}

 and

\vspace{2mm}

{\bf Hendrik B. Geyer}

\vspace{2mm}

{\it
Institute of Theoretical Physics, University of Stellenbosch,

Matieland 7602, South Africa

e-mail: hbg@sun.ac.za
}

\vspace{3mm}


\end{center}

\vspace{3mm}

\section*{Abstract}

For a specific exactly solvable $2\times 2$ matrix model with a ${\cal PT}$-symmetric Hamiltonian
possessing a real spectrum, we construct {\em all} the eligible
physical metrics ${\Theta}>0$ and show that {\em none} of them
admits a factorization $\Theta={\cal CP}$ in terms of an
involutive charge operator ${\cal C}$. Alternative ways of restricting
the physical metric to a unique form are briefly discussed.

\vspace{5mm}

\noindent
{\it PACS:} 03.65.-w; 03.65.Ca; 03.65.Ta\\
{\it Keywords:} Quasi-Hermitian, non-Hermitian, PT-symmetric quantum mechanics; metric; charge operator
%

\newpage

\section{Introduction}

The recent perceivable growth of interest in quasi-Hermitian
Hamiltonians $H \neq H^\dagger$ with real spectra (cf. the Appendix
for a brief description) has, amongst other reasons, received an impetus from their
possible deep relevance in field theory \cite{BB}.
Using several numerical methods the authors of the latter pioneering letter
demonstrated that the family of the manifestly non-Hermitian
Hamiltonians
 \be
 H = H(\nu)= -\frac{\hbar^2}{2m^2}\frac{d^2}{dx^2} + x^2({\rm i}x)^\nu,
 \ \ \ \ \ \ \nu \geq 0
 \label{SEb}
 \ee
possess (only) real energies $E_n(\nu)$ which smoothly vary with
the exponent $\nu>0$. In the limit $\nu \to 0^+$ this spectrum coincides with
the well-known harmonic oscillator equidistant set $E_n(0)=2n+1$,
$n = 0, 1, \ldots$ in units where $\hbar = 2m = 1$.

Although these observations have only been supported by
rigorous proofs a few years later \cite{DDT}, the appealing
contrast between the manifest {\em non}-Hermiticity of the Hamiltonian
(\ref{SEb}) and the strict absence of {\em any} instability of the
levels (${\rm Im}\,E_n = 0$ for {all} $n$) proved inspiring
\cite{all}. Bender and Boettcher's hypothesis that the
manifest ${\cal PT}$-symmetry, $H (\nu){\cal PT} = {\cal PT}H(\nu)$,
of the Hamiltonian in question (where ${\cal P}$ indicates parity and
${\cal T}$ is complex conjugation \cite{BBjmp}) underpins this property, has been
confirmed by the strict reality of the spectrum in a number of
exactly solvable examples \cite{ES}. Moreover, Mostafazadeh
\cite{ali} then broadened the range of the hypothesis by pointing out
that Hamiltonians with  similar properties may be identified
among all the pseudo-Hermitian operators which appear Hermitian in
a suitable indefinite (pseudo)metric generalization~$\eta$ of
parity,
 \be
 H^\dagger  = \eta \,H\,\eta^{-1}, \ \ \ \
  \ \ \ \ \ \eta=\eta^\dagger\,.
 \label{pseudohe}
 \ee
Alternatively, in the light of the recent studies \cite{couchan}
one may select the ``generalized parity" $\eta$ in
eq.~(\ref{pseudohe}) as a non-Hermitian operator, provided only
that the product ${\cal S}= \left [\eta^{-1}\right ]^\dagger\eta
\neq I$ remains a symmetry of our Hamiltonian, $H {\cal S} =
{\cal S} H$.

Coming full circle, the focus then returned to a re-discovery of the
well-known fact that the compatibility of all similar models
with the postulates of Quantum Mechanics necessitates that {\em
besides} the property (\ref{pseudohe}), {\em all} the models in
question must {\em also} satisfy  {\em another, independent} condition,
termed quasi-Hermiticity \cite{SGH}:
 \be
 H^\dagger  = \Theta \,H\,\Theta^{-1}, \ \ \ \ \ \ \ \ \
  \Theta=\Theta^\dagger>0\,.
 \label{quasihe}
 \ee
The symbol $\Theta$ for the positive definite metric here replaces $T$,
originally introduced in the paper~\cite{SGH} (where the terminology {\it quasi-Hermitian} was coined),
and which may now be confusing in view of the central role of the time-reversal operator
${\cal T}$ in ${\cal PT}$-symmetric quantum mechanics. In
related more recent literature one also finds some alternative symbols, e.g.\
$\eta_+$ \cite{ali}, ${\cal CP}$ \cite{BBJ}, $U$ \cite{Jones},
${\cal PQ}$ \cite{ja}, all of them referring to
the metric in the Hilbert space ${\cal H}$ of physical
states.

\section{ Ambiguity of the metric \label{twodim} }

One has to note and emphasize that the specification of the
operator $\Theta$ from the Hamiltonian does not determine the metric uniquely \cite{SGH}.
Indeed, once we fix $H$, the linear eq.~(\ref{quasihe}) may
lead to many different positive and Hermitian solutions
$\Theta$ (\cite{alib}, cf. Appendix). The existing approaches to
determine a unique $\Theta$ from subsidiary conditions include

\begin{itemize}

\item
a systematic introduction of further ``external" observables
to complete, with the Hamiltonian, an irreducible set of observables.
Requiring the associated operators $A_1$,
$A_2$, $\ldots$ to be quasi-Hermitian with respect to the {\em same} metric
$\Theta$, then completely eliminates any remaining freedom in
$\Theta$  \cite{SGH};

\item
a requirement that the metric $\Theta$ be factorized using a
``charge" operator $\cal C$: ${\cal C}\equiv \Theta\,{\cal P}^{-1}$
\cite{BBJb}, or
``quasi-parity" ${\cal Q}\equiv {\cal P}^{-1}\, \Theta$
\cite{pseudo}. Then, the key idea of the reduction of the freedom
in $\Theta $
relies on the ``natural" involution properties ${\cal C}^2=I$
and ${\cal Q}^2=I$,  respectively;

\item
a separable representation of $\Theta$ which enables one to fix
the free constants in each term separately \cite{Batal};

\item
a transition to the partial-differential-equation re-arrangement
of the quasi-Hermiticity condition (\ref{quasihe}) which reduces
the ambiguity of $\Theta$ by the specification of the related
boundary conditions \cite{Scholtz}.

\end{itemize}

\noindent Some of these alternatives are discussed here via an
explicit analysis of a schematic model.

\subsection{A $2\times 2$ matrix example }

Consider an harmonic oscillator basis in which parity is explicitly indicated,
$\{ \,|n,\pm\rangle \,\}$. (Here $n = 0,
1, \ldots$ with parity $+$ or $-$). Generically \cite{humor}, Hamiltonians $H(\nu)$ of the complex symmetric type (\ref{SEb})
will take the schematic infinite-dimensional real, but {\em non-symmetric}, form
  \be
 H (\nu) = \left (
 \begin{array}{cc}
 S&B\\
 -B^T&L
 \ea
 \right )\,,
 \label{neherma}
  \ee
when represented in a basis which uses normalized states of the form $|n,+\rangle$ and $i|n^\prime,-\rangle$,
with $S$ and $L$ symmetric, and $B$ not necessarily so.  (A specific example is the well studied imaginary potential
$V(x) = ix^3$ which has non-zero matrix elemenets only between  basis states with different parity;
these matrix elements are then real with our choice of basis.)

The usual hermiticity property of Hamiltonians is here replaced
by a matrix version of the pseudo-Hermiticity of $H$ { with
respect to the indefinite (pseudo)metric} {\em matrix} $P$,
 \be
 \left [
 H(\nu)
 \right ]^\dagger = P\,
 H(\nu)\,P^{-1},
  \ \ \ \ \ \ \ \ \ \ \
 P = P^{-1}=\left (
 \begin{array}{cc}
 I&0\\
 0&-I
 \ea
 \right )=P^\dagger.
 \label{pseudoh}
  \ee
A ``better" basis might have been chosen such that both the
matrices $S$ and $L$ become diagonal.

In order to clarify the essence of the model (\ref{neherma}) let
us contemplate a violation of the parity by such an {\it ad hoc}
potential which leaves just a single matrix element in the matrix
$B$ of eq.~(\ref{neherma}) different from zero, $B_{IJ} \neq 0$.
In such a situation (with diagonal $S$ and $L$) the coupling is
only introduced between the $I-$th even energy $S_{II}$ and the
$J-$th odd energy $L_{JJ}$. Hence, the solution of the full
Schr\"{o}dinger equation degenerates to the analysis of the
two-dimensional matrix problem
 \be
 H
  \left (
  \ba
  x\\
  y
  \ea
  \right ) = E\,
  \left (
  \ba
  x\\
  y
  \ea
  \right ), \ \ \ \
 H =
 \left (
 \begin{array}{cc}
 S_{II}&B_{IJ}\\
 -B_{IJ}&L_{JJ}
 \ea
 \right )\,.
 \label{toyes}
 \ee
In contrast to our previous considerations, we have to deal here
with only three real matrix elements.

We emphasise that the ${\cal PT}$-symmetric matrix Hamiltonian (\ref{toyes}) could also have
been written down without any reference to the Hamiltonian (\ref{SEb}), the present link only
serving as additional motivation.

\subsection{The metric $\Theta$}

For the sake of simplicity we may drop the subscripts and shift
the origin of the energy scale in eq.~(\ref{toyes}),
 \be
 H=
 \left (
 \begin{array}{cc}
 -{D}&B\\
 -B&{D}
 \ea
 \right ),
  \ \ \ \ \ \ \ \ \ \ \
 {\cal P} = {\cal P}^{-1}=\left (
 \begin{array}{cc}
 1&0\\

 0&-1
 \ea
 \right )={\cal P}^\dagger.
 \label{hamoun2}
 \ee
Remember that
our first requirement is that all the eigenvalues, viz, the doublet
 \ben
 E_\pm = \pm \sqrt{{D}^2-B^2}
 \een
remain real, i.e.,
 \ben
 B = {D}\,\cos \alpha, \ \ \ \ E_\pm = \pm {D}\,\sin \alpha,
 \ \ \ \ \ \alpha \in (0,\pi)\,.
 \een
We have to exclude both the endpoints $\alpha = 0, \pi$ where one
encounters the exceptional point (i.e., the geometric and
algebraic multiplicities of $E=0$ become different there). Also
the point $\alpha = \pi/2$ is not interesting since our
Hamiltonian becomes diagonal and Hermitian there. Finally, we may
choose any overall factor ${D}$ and  assume that $\alpha \in (0,
\pi/2)$ without any loss of insight mediated by the model.
Equation (\ref{hamoun2}) becomes replaced by its still fully
representative one-parametric ${D}=1$ version
 \be
 H=
 \left (
 \begin{array}{cc}
 -1&\cos \alpha\\
 -\cos \alpha&1
 \ea
 \right ).
 \label{hamounovich2}
 \ee
To this Hamiltonian we now have to assign a Hermitian
metric operator containing four real parameters in general,
  \ben
  \Theta = \left (
 \begin{array}{cc}
 a&b+ic\\
 b-ic&d
 \ea
 \right ).
  \een
This operator must satisfy eq.~(\ref{quasihe}), $\Theta H=H^T
\Theta$. Insertion of this general form of $\Theta$ shows that
the quasi-hermiticity condition (\ref{quasihe}) implies two restrictions
on the parameters of the metric:
\be
 2b=-(a+d)\cos \alpha; \quad c=0\; .
 \ee
We must furthermore demand that $b \neq 0 \neq a+d = 2Z$, otherwise the metric could not be positive (or negative)
definite as required. In this notation we may put $a=Z(1+\xi)$ and
$d=Z(1-\xi)$ with any real $\xi$.

The scale factor $Z$ is again arbitrary and may be set equal to
one. In this way we arrive at the most general solution of our
present $N=2$ version of eq.~(\ref{quasihe}),
 \be
 \Theta =
 \left (
 \begin{array}{cc}
 1+\xi&-\cos \alpha\\
 -\cos \alpha&1-\xi
 \ea
 \right )\,.
 \label{generalm}
  \ee
For such a two-dimensional matrix family both the eigenvalues are
available in closed form,
 \ben
  \theta_\pm = 1 \pm \sqrt{\xi^2+\cos^2 \alpha}\; .
 \een
It is simple to conclude that {\em both} of them remain
non-degenerate and positive if and only if
 \be
    1>\sqrt{\xi^2+\cos^2\alpha}>0\; .
    \label{finalcond}
    \ee
This means that we may set
 \be
 \xi = \sin \alpha  \sin \gamma,
\ \ \ \gamma \in [0, \pi/2)
  \label{finalre}
 \ee
with one independent free real parameter
$\gamma$. This completes our construction of the general metric $\Theta$.

\section{Viable restrictions to fix the metric  \label{konec} }

Although all the so-called ${\cal PT}-$symmetric Hamiltonians $H
\neq H^\dagger$ with real spectra can be treated as Hermitian with
respect to many nontrivial {\it ad hoc} metrics $\Theta \neq I$,
the choice of restrictions to obtain a unique ``physical'' metric $\Theta_{\rm phys}$
remains largely unchartered. Several aspects of this open problem may be
discussed via our highly schematic two-state model.

A real and symmetric $\Theta$ may also be generated directly from the
bi-orthogonal set of eigenstates of $H$ whenever a pseudo-Hermitian
Hamltonian is diagonalizable in a suitable bi-orthogonal
basis (see Ref.\ (\cite{ali}, and also eq.~(\ref{representace}) in the
Appendix below).

\subsection{Establishing an irreducible set of observables }

As we already mentioned, there have been several proposals to reduce the
freedom in the choice of $\Theta$ which may be chosen as {\em any}
solution of the operator eq.~(\ref{quasihe}). In particular, the
authors of Ref.\ \cite{SGH} showed that one has
{\em to choose} a set of irreducible or ``natural'' observables ${\cal A}_j$ {\em
and to demand} that
 \be
 \left [
 {\cal A}_j
 \right ]^\dagger = \Theta\,
 {\cal A}_j\,\Theta^{-1}
 \label{comuquasih}
  \ee
for {\em all} of them. However, different choices will lead to different
quantum mechanical frameworks, and there is no general algorithm whereby the set may be
completed after an initial choice of an observable (or observables) has been made to supplement the Hamiltonian.
Under such a scenario let us assume that
the ``natural" observables ${\cal A}_j$ will mimic the
``irreducible" set of coordinates $x$ and momenta $p$ (as mentioned
in ref.~\cite{SGH}) by being represented by the {\em Hermitian}
matrices,
 \ben
 {\cal A}_j =\left (
 \begin{array}{cc}
 A_j&B_j\\
 B_j&D_j
 \ea
 \right )\,.
  \een
Such a family of the observables, characterized, {\it a priori}, by the
three real parameters $A$, $B$ and $D$ must remain compatible with the
original quasi-Hermiticity constraint~(\ref{comuquasih}).
This represents (for all possible $j$) the constraint
\ben
 2\xi B = (D-A) \cos \alpha.
 \een
As the choice of $B=0$ would imply that ${\cal
A} \sim I$ are trivial, we are allowed to assume that $B\neq 0$.
In this way the choice of a single ${\cal A}_1={\cal A}$ fixes the
freedom in our physical metric operator completely,
 \ben
 \xi_{\rm phys} = \frac{D-A}{2B}\,\cos \alpha \; .
 \een
We may conclude that the approach proposed in ref.~\cite{SGH}
fixed the parameter in our toy metric and leads to
unique ``physics".

It is instructive to notice that for the latter purpose one,
and only one, {\em general} auxiliary observable ${\cal A}_1$
proved sufficient in our schematic model. It will be interesting to
find some other and, hopefully, less schematic models with this
property, which seems to be reminiscent of one-dimensional
standard quantum mechanics where $x$ and $p$ form an irreducible set, the Hamiltonian
generally being an auxiliary or derived operator, $H=H(x,p)$.

\subsection{Factorization in terms of a charge operator}

Our final comment concerns the connection of our example with the
popular postulate of the factorization $\Theta = {\cal CP}$
accompanied by the requirement that the factor ${\cal C}$ may be
interpreted as an operator of a ``charge" with the property $
{\cal C}^2 = I$ (cf., e.g., ref. \cite{BBJ}. In our example we may
easily derive the explicit formula
 \ben
 {\cal C}^2=
  \left (
 \begin{array}{cc}
 (1+\xi)^2-\cos^2\alpha & 2\xi\cos \alpha\\
 -2\xi\cos \alpha & (1-\xi)^2-\cos^2\alpha
 \ea
 \right ).
 \een
Obviously, the requirement $ {\cal C}^2 = I$ implies that $\xi=0$.
This forces us to set $\cos \alpha = 0$. We see
that this in fact eliminates {\em all} the
nontrivial Hamiltonians in our family. In other words, for all
the non-diagonal non-Hermitian models with $\alpha \neq \pi/2$ the
{\em involutive} charge operator {does not exist at all.}

\section{Conclusion}

We have motivated and studied a particular $2\times 2$ ${\cal PT}$-symmetric matrix Hamiltonian,
characterizing and constructing from various points of view a {\em unique}
positive definite metric which would render the Hamiltonian quasi-Hermitian, and thus amenable to a
standard quantum mechanical interpretation.

We conclude that while it is possible to find such metrics, the insistence on factorisation,
$\Theta={\cal CP}$, which is usually enforced in ${\cal PT}$-symmetric quantum mechanics,
is inapplicable here. Note in this respect that the Hamiltonian matrix (\ref{neherma}) is
non-symmetric, in contrast to the $2\times 2$ model studied by Bender {\it et al} \cite{BBJb} for which a $\cal C$-operator had in fact been constructed.
In general further work is called for to elucidate the construction and implications of various
approaches to identify and implement a unique metric in quasi-Hermitian quantum mechanics.

\vspace{5mm}

\section*{Acknowledgement}

MZ (partially supported by the IRP AV0Z10480505
and by the project LC06002
of the Ministry of Education, Youth and Sports of the Czech
Republic)
gratefully acknowledges the
inspiring atmosphere and the hospitality of the Physics Department
of the University of Stellenbosch where this work has been mainly
performed.


\newpage

\section*{Appendix: Metrics $\Theta\neq I$ in
Hilbert space }

According to the standard postulates of Quantum Mechanics, the
states of a given system may be represented by elements $|\psi
\rangle$ of a Hilbert space ${\cal H}$, endowed by a positive
definite  metric $\Theta$. The statement about the metric is mostly omitted,
as the standard metric $\Theta = I$ in $L^2(-\infty,\infty)$ is assumed.
A system is then fully characterized
by a set of its observable characteristics, e.g.\ by the
energy $E$ and by some other real, measurable quantities $a_i$, $i
= 2,3, \ldots, N^{({\rm obs})}$, expected to lie in some respective
subsets ${\cal D}_i$ of $I\!\!R$.

As pointed out in Ref.\ \cite{SGH} the framework of Quantum Mechnics remains intact
when a positive definite metric $\Theta \neq I$ is introduced.
In such a context the quantum description of any system requires that
its observables $a_i$ are represented by the operators $A_i$ in
${\cal H}$ which are Hermitian with respect to the ``physical"
metric  $\Theta$,
 \be
 A_i^\dagger = \Theta\,A_i\,\Theta^{-1}.
 \label{qua}
 \ee
Such operators, including $H$, have been termed ``quasi-Hermitian" in Ref.\ \cite{SGH}.

The concept re-emerged in the framework of ${\cal PT}-$symmetric
Quantum Mechanics \cite{BB} where $H \neq H^\dagger$, so that,
typically, one must solve not only the ``direct" Schr\"{o}dinger
bound-state problem, but also its Hermitian conjugate partner,
 \be
 H \,|n\rangle = E_n\,|n\rangle, \ \ \ \ \ \
 \langle\langle n|\,H=\langle\langle n|\,E_n\,.
 \ee
Here we employed the matrix-algebra-inspired ``left-action"
convention  and introduced a double bra symbol
$\langle\langle n|$ for the conjugate eigenkets of $H^\dagger$.

Focussing on the Hamiltonian $H \neq
H^\dagger$ only, the spectrum may be degenerate (and/or complex where no $\Theta > 0$ exists), while
the related set of wave functions may also prove incomplete in
general. All such degeneracies will be skipped and avoided in this short exposition.
Thus, we may assume the biorthogonality
($\langle\langle n| m \rangle = 0$ iff $m\neq n$)
and completeness of our bound states,
 as well as the existence of a  metric $\Theta \neq I$ in
the Hilbert space, rendering the Hamiltonian quasi-Hermitian,
 \be
 H^\dagger = {\Theta}\,H\,{\Theta}^{-1}\,, \ \ \ \ \
  \ \ {\Theta} = {\Theta}^\dagger >0 \; .
 \label{uquasih}
 \ee
(See Ref.~\cite{SGH} for more details.) The insertion of the
formal spectral expansion
 \be
 H = \sum_{n=0}^{\infty}\,|n\rangle\,
 \frac{1}{\langle\langle n|n\rangle}\,
 \langle\langle n|
 \label{spectralH}
 \ee
into eq.~(\ref{uquasih}) indicates that
 \be
 \Theta = \sum_{n,m}\,|n\rangle\rangle\,s_{n,m}\,\langle\langle m|\,.
 \label{representace}
  \ee
Inserting the spectral form (\ref{spectralH}) and this ansatz into the defining eq.\ (\ref{uquasih}),
 leads to the conclusion
that the array of coefficients must remain diagonal,
$s_{n,m}=\delta_{n,m}\,s_m$,
 \be
 \Theta = \sum_{m}\,|m\rangle\rangle\,s_{m}\,\langle\langle m|\,.
 \label{dorepresentace}
  \ee
These coefficients must be real (due to the Hermiticity
requirement ${\Theta} = {\Theta}^\dagger$) and positive (in order
to guarantee the positive-definiteness of ${\Theta} >0$). This presents
an explicit construction of the metric $\Theta$, the
remaining freedom in $s_m$ of course again reflecting the fact that
quasi-Hermiticity of the Hamiltonian alone does not determine the metric uniquely.




\newpage

\begin{thebibliography}{00}

\bibitem{BB}
C. M. Bender and S. Boettcher, Phys. Rev. Lett. { 80} (1998) 4243.

\bibitem{DDT}
P. Dorey, C. Dunning and R. Tateo, J. Phys. A: Math. Gen. 34
(2001) 5679;

K. C. Shin, Commun. Math. Phys. 229 (2002) 543.

\bibitem{all}
F. M. Fern\'{a}ndez, R. Guardiola, J. Ros and M. Znojil, J. Phys.
A: Math. Gen. 31 (1998) 10105;

B. Bagchi, S. Mallik and C. Quesne, Mod. Phys. Lett. A 17 (2002)
1651;

A. Mostafazadeh,
Class. Quantum Grav. 20 (2003) 155;

H. Langer and Ch. Tretter, Czech. J. Phys. 54 (2004) 1113;

Czech. J. Phys. 55 (2005) 1045 - 1192;
issue dedicated to {\it Pseudo-Hermitian Hamiltonians
in Quantum Mechanics}.

\bibitem{BBjmp}
C. M. Bender, S. Boettcher and P. N. Meisinger, J. Math. Phys. 40
(1999) 2201.

\bibitem{ES}
A. Andrianov, F. Cannata, J.-P. Dedonder and M. V. Ioffe,
Int. J. Mod. Phys. A 14 (1999) 2675;

M. Znojil,
Phys. Lett. A 259 (1999) 220;


G. L\'evai and M. Znojil,
J. Phys. A: Math. Gen. 33(2000) 7165;



B. Bagchi and C. Quesne,
         Phys. Lett. A 273 (2000)
                285;

M. Znojil, Phys. Lett. A. 285 (2001) 7;


M. Znojil and M. Tater,
J. Phys. A: Math. Gen. 34 (2001) 1793;


S. Albeverio, S.-M. Fei and P. Kurasov, Lett. Math. Phys. 59
(2002) 227;

V. Jakubsk\'{y}, Czech. J. Phys. 54 (2004) 67;

A. Sinha and P. Roy, Czech. J. Phys. 54 (2004) 129;

H. B\'{\i}la, V. Jakubsk\'{y}, M. Znojil, B. Bagchi, S. Mallik and
C. Quesne,
Czech. J. Phys. 55 (2005) 1075.














\bibitem{ali}
A. Mostafazadeh, J. Math. Phys. 43 (2002) 205 and 2814.




\bibitem{couchan}
M. Znojil, J. Phys. A: Math. Gen. 39 (2006) 4047;

M. Znojil, Phys. Lett. A (2006), to appear (quant-ph/0601048).

\bibitem{SGH}
F. G. Scholtz, H. B. Geyer and F. J. W. Hahne, Ann. Phys. (NY) 213
(1992) 74.


\bibitem{BBJ}
C. M. Bender, D. C. Brody and H. F. Jones, Phys. Rev. Lett. 89
(2002) 0270401;
%
%
%

\bibitem{Jones}
H. F. Jones, J. Phys. A: Math. Gen. 38 (2005) 1741.

\bibitem{ja}
M. Znojil,

in
``Symmetry Methods in Physics",
Ed. C. Burdik, O. Navratil and S. Posta,
 JINR, Dubna, 2004 (CD ROM, also hep-th/0408081).

\bibitem{alib}
A. Mostafazadeh,
 J. Phys. A: Math. Gen. 38 (2005) 6557;
 erratum, {\it ibid} 8185;

A. Mostafazadeh, quant-ph/0508195.

\bibitem{BBJb}
C. M. Bender, D. C. Brody, H. F. Jones,
Phys. Rev. Lett. 93 (2004) 251601;
%



 C. M. Bender, D. C. Brody and H. F. Jones,
 Phys. Rev. D 70 (2004) 025001; Erratum-ibid. D 71 (2005) 049901;
%



C. M. Bender and  H. F. Jones,
Phys. Lett. A 328 (2004) 102;

C. M. Bender, S. F. Brandt, J-H. Chen and Q. Wang,
Phys. Rev. D 71 (2005) 065010.


\bibitem{pseudo}
M. Znojil, Rendiconti del Circ. Mat. di Palermo, Ser. II, Suppl.
72 (2004) 211 (math-ph/0104012);

M. Znojil,
J. Phys. A: Math. Gen. 39 (2006) 441.


\bibitem{Batal}
A. Mostafazadeh and A. Batal, J. Phys. A: Math. Gen. 37 (2004)
11645.




\bibitem{Scholtz}
F. G. Scholtz and H. B. Geyer, Phys. Lett. B 634 (2006) 84.


\bibitem{humor}
M. Znojil, math-ph/0106021;

M. Znojil, J. Nonlin. Mat. Phys. 9, Suppl. 2 (2002) 122.


\end{thebibliography}
\end{document}